\documentclass[12pt,preprint]{aastex}

\begin{document}

\title{Deep Infrared Imaging of the Microquasars 1E1740-2942 and GRS
1758-258\footnote{Based in part on observations obtained at the
W.M. Keck Observatory, which is operated by the California Association
for Research in Astronomy, a scientific partnership among the
California Institute of Technology, the University of California, and
the National Aeronautics and Space Administration.}}

\author{S.S. Eikenberry\altaffilmark{1,2},
W. J. Fischer\altaffilmark{1,3}, E. Egami\altaffilmark{2},
S.G. Djorgovski\altaffilmark{4}}

\altaffiltext{1}{Astronomy Department, Cornell University, Ithaca, NY
14853} \altaffiltext{2}{Physics Department, California Institute of
Technology, Pasadena, CA 91125} \altaffiltext{3}{Ritter Astrophysical
Research Center, University of Toledo, Toledo, OH 43606}
\altaffiltext{4}{Astronomy Department, California Institute of
Technology, Pasadena, CA 91125}

\begin{abstract}

	We present deep infrared ($2.2 \mu$m) imaging of the Galactic
microquasars 1E1740-2942 and GRS 1758-258 using the Keck-I 10-meter
telescope in June 1998.  The observations were taken under excellent
seeing conditions ($\sim 0.45 \arcsec$ full-width half-maximum), making
them exceptionally deep for these crowded fields.  We used the
USNO-A2.0 catalog to astrometrically calibrate the infrared images
(along with an optical CCD image in the case of GRS 1758-258),
providing independent frame ties to the known radio positions of the
objects.  For 1E1740-2942, we confirm potential candidates for the
microquasar previously identified by Marti et al., and show that none
of the objects near the microquasar have varied significantly from
1998 to 1999.  For GRS 1758-258, our astrometry indicates a position
shifted from previous reports of candidates for the microquasar.  We
find no candidates inside our 90\% confidence radius to a $2 \sigma$
limiting magnitude of $K_s = 20.3$ mag.  We discuss the implications
of these results for the nature of the microquasar binary systems.

\end{abstract}


\section{Introduction}

	The discovery of Galactic relativistic jet systems -- the
``microquasars'' [\citep{Felix94}; \citep{Tingay}] -- promises to
revolutionize the study of jets in black hole systems.  Due to their
proximity and smaller mass, these X-ray binaries reveal
smaller/fainter features and vary more rapidly than their
extragalactic cousins, the quasars.  Thus, patterns of variability
which repeat on timescales of decades to centuries for the quasars can
be observed on timescales of minutes to days in their Galactic
counterparts.  This property makes the microquasars excellent test
laboratories for investigating the black-hole/relativistic-jet
connection.

	Based on their X-ray and radio properties, 1E1740-2942 and GRS
1758-258 are both considered strong microquasar candidates (in fact,
the term ``microquasar'' was first used to describe 1E1740-2942).  They
are the two brightest hard X-ray sources near the Galactic Center at
energies $>50$ keV, and their hard X-ray tails make them likely black
hole candidates \citep{Sunyaev}.  Furthermore both objects display
radio lobes, indicating the presence of collimated jet outflows
[\citep{Felix92}; \citep{Luis}].  In their hard X-ray spectra and
variability, both objects closely resemble the black hole candidate
Cyg X-1 (e.g. \citet{Kuz}), suggesting that they may be
black-hole/O-star binaries.  Both objects lie in the Galactic Plane
with high optical obscuration ($A_V \sim 10$ for GRS 1758-258 and $A_V
\sim 25-50$ for 1E1740-2942), so IR searches for companions/counterparts
have been performed.  However, initial searches have returned negative
results, with upper limits of $K>17$ (e.g. \citet{George}), and
\citet{Felix91} failed to detect the expected radio emission from the
compact HII region an O-star would form.

	Recently, faint IR candidate counterparts for both systems
have been reported.  For GRS 1758-258, \citet{Marti98} have found 3
obscured optical stars in or near the radio error circle.  Similarly,
\citet{Marti00} have reported 1 or more faint IR stars inside the
error circle for 1E1740-2942.  However, due to the crowded nature of the
fields, none of the candidates can be singled out as a certain
microquasar counterpart.

	In order to identify such counterparts, we obtained deep
$K_s$-band images of both microquasar fields using the Keck 10-m
telescope in June 1998.  In Section 2, we describe the observations
and data reduction we performed.  In Section 3, we present our
astrometric solutions for the locations of the microquasars on the IR
images.  In Section 4, we present photometry of the candidate
counterparts and limits on potential undetected counterparts.  In
Section 5, we discuss the implications of these results, and in
Section 6 we present our conclusions.

\section{Observations and Data Reduction}

\subsection{Keck IR Observations}

\subsubsection{June 1998}

	We observed the fields of both microquasars on June 1-2, 1998
UTC using the Keck Observatory's Near-Infrared Camera (NIRC --
\citet{NIRC}) on the Keck I telescope.  NIRC has a 256x256-pixel InSb
array with a 0.15 $\arcsec/{\rm pixel}$ plate scale, giving a $38
\arcsec$ field of view.  Due to the high extinction expected given the
X-ray absorption towards these objects, we used a $K_s$ filter,
centered at $2.15 \mu$m.  Conditions were not perfectly photometric on
either night, with high thin cirrus clouds present, and monitoring of
standard stars throughout the observing run indicated that the
atmospheric absorption losses were variable at the $\sim 10\%$ level.
The seeing conditions were exceptionally good, ranging from $\sim 0.35
\arcsec$ to $\sim 0.55 \arcsec$, with extended periods of seeing $<
0.5 \arcsec$.

	For both objects, we used a random, non-repeating dither
pattern, moving the telescope in $10 \arcsec$ steps on the sky around
the radio positions of the microquasars.  We took 12 coadded 5-second
exposures at each position, for a total of 60 seconds on-source per
position.  We repeated this dithering for a total of 94 minutes of
exposure on GRS 1758-258 on June 1 and 99 minutes on June 2.  For
1E1740-2942, we obtained 104 minutes of total exposure on each night.

	We created a sky background frame by median-combining the
dithered images for each object and each night separately.  We then
subtracted the sky background from each frame, divided by a flatfield
created from the dithered sky frame, and interpolated over bad pixels
and cosmic ray events.  For GRS 1758-258, we then measured the
centroid position of several bright stars in each frame and used them
to determine its shift relative to a reference frame.  We then shifted
the frames accordingly and averaged them to obtain a master image of
the field of GRS 1758-258 (see Figure 1).

	We treated the images of 1E1740-2942 similarly up to the
shifting and adding stage.  Here the extremely crowded nature of the
1E1740-2942 field necessitated a different procedure.  Rather than
simply shifting and adding the images, we first measured the
full-width half-maximum (FWHM) of the stellar point spread function
(PSF) using several bright stars in the field.  We then combined the
images as above, but rejecting any images with FWHM greater than
certain values.  We tried several trial values for the FWHM cut, and
found that $0.45 \arcsec$ FWHM provided the best compromise between
good resolution (low FWHM) and total exposure time (high FWHM).  We
present the resulting master image of 1E1740-2942 in Figure 2.

	We also observed the UKIRT photometric standard star FS28.  We
took five dithered images of this star, which we used to create a sky
frame.  We subtracted the sky frame, divided by a flatfield, and
interpolated over bad pixels and cosmic rays.

\subsubsection{April 1994}

	We also observed 1E1740-2942 using NIRC on Keck I on 5 April
1994.  We operated in a very similar mode to that described above, but
with a total integration time of 720 seconds.  While conditions were
photometric on this night, the seeing was rather inferior ($\sim 1.0
\arcsec$ FWHM in $K_s$), providing limited utility in this crowded
field.

\subsection{CCD Observations}

	In order to accurately calibrate the astrometric reference
frame of the IR images of GRS 1758-258, we obtained wide-field ($8
\arcmin \times 8 \arcmin$) CCD images of its field in the I band on
June 29, 1998 using the Palomar 60-inch telescope.  The total exposure
time in each band was 300 seconds, and the typical stellar PSF had
FWHM $\sim 1.2 \arcsec$.  We subtracted the CCD bias from each frame,
divided by a flatfield, and interpolated over bad pixels and cosmic
rays.

\section{Astrometric Calibration of IR Images}

\subsection{1E1740-2942}

	In the case of 1E1740-2942, we identified 4 stars in Figure 2
which are listed in the USNO-A2.0 astrometric catalog \citep{USNO}.
One of them is a close double, which complicates the astrometry
sufficiently that we dropped it from our analysis.  We used the
remaining 3 stars to determine the ``best-fit'' astrometric solution
for Figure 2 and locate the radio position for 1E1740-2942 on Figure 2
(marked by an ``X'').  We use the VLA position $\alpha_{\rm J2000} =
17^h \ 43^m \ 54^s.83$ and $\delta_{\rm J2000} = -29^o \ 44
\arcmin \ 42 \arcsec .60$ \citep{Marti00}.  We then calculated the RMS
residuals for the USNO star positions from this solution, which we
find to be $\pm 0.37 \arcsec$.  Note that this includes the internal
uncertainties in the USNO positions, which are typically $\sim 0.25
\arcsec$.  Including the $1 \sigma$ uncertainties in the radio
position ($0.1 \arcsec$ -- \citet{Luis}) and in the radio-optical
frame tie ($< 0.1 \arcsec$ -- \citet{daSilva}), we arrive at a 90\%
positional uncertainty of $\pm 0.65 \arcsec$.  Figure 3 shows a
section of Figure 2 close to the position of 1E1740-2942, including
the 90\% error circle.  Comparison of this position to similar results
obtained independently by \citet{Marti00} show a close match between
them, to better than $\sim 0.2 \arcsec$.

\subsection{GRS 1758-258}

	In the case of GRS 1758-258, we could not identify any
suitable USNO-A2.0 stars in the field of Figure 1, and thus turned to
the larger-field CCD image as a secondary astrometric reference.  We
identified 12 USNO stars within $\sim 1.5 \arcmin$ of the approximate
radio position for GRS 1758-258.  We use the VLA position $\alpha_{\rm
J2000} = 18^h \ 01^m \ 12^s.395$ and $\delta_{\rm J2000} = -25^o \ 44
\arcmin \ 35 \arcsec .90$ \citep{Felix93}.  We then obtained an
astrometric solution for the CCD image in this region using these
stars.  We then transferred these coordinates onto 5 stars near the
position of GRS 1758-258 which were visible in both the I-band CCD
image and the K-band NIRC image, and used them to derive a secondary
astrometric solution for the NIRC image.  The resulting position is
indicated on Figure 1 by an ``X''.  The RMS residuals to the CCD
astrometric solution were $0.22 \arcsec$, and the RMS residuals of the
IR stars to the CCD solution were $\sim 0.06 \arcsec$.  Including the
$0.1 \arcsec$ $1 \sigma$ uncertainties in both the radio position of
GRS 1758-258 \citep{Felix93} and the optical-radio frame tie
\citep{daSilva}, we obtain a 90\% positional uncertainty of $0.50
\arcsec$ for our position.  Figure 4 shows a section of Figure 1 close
to the position of GRS 1758-258, including the 90\% error circle.
Comparison of this position to similar results obtained independently
by \citet{Marti98} show a $\sim 1.5 \arcsec$ offset between the two
positions, which is large enough that the two 90\% error circles just
barely touch.  The approximate location of the Marti et
al. astrometric solution for GRS 1758-258 is also indicated in Figure
4.

\section{Photometry of Candidates and Limits on Counterparts}

\subsection{1E1740-2942}

	In Figure 3, we can see several stars near the position of
1E1740-2942, which we have identified by the numbering system introduced
by \citet{Marti00}.  We do not detect any objects other than the stars
identified by Marti et al., nor do we fail to detect any objects
identified by Marti et al.  In addition, with our improved $0.45
\arcsec$ angular resolution, we confirm the suggestion of
\citet{Marti00} that ``Star 4'' is an extended object.  However, we
cannot determine whether the extended appearance is due to truly
diffuse emission or simply the overlap of several stellar images in
this crowded field.

	In Table 1, we present photometric results for the numbered
stars in Figure 3, using FS28 as the photometric standard.  This star
has $m_{K} = 10.597$ mag and $m_H = 10.644$ mag, based on which we
estimate $m_{Ks} = 10.62 \pm 0.03$ mag.  Comparing the PSF-weighted
photometry of the 5 images of FS28, we find an RMS variation of $\pm
0.02$ mag in the measured flux levels.  We performed photometry of the
stars in the 1E1740-2942 field using the same PSF-weighted algorithm and
assuming an average value for the sky background in the field.  Due to
the extreme crowding of the field, we find significant gradients in
the apparent background level in this region of the field, and we
estimate our uncertainty in the background level per pixel to be 20\%.
In addition, we include an estimated 10\% uncertainty in the
photometry due to transparency variations on these non-photometric
nights.

	All of the photometric results for the stars in Table 1 match
the photometry of \citet{Marti00} within the uncertainties, indicating
that none of them varied significantly between June 1998 and May 1999.
If we consider the exact center of our astrometric error circle for
1E1740-2942, we can place an upper limit on the presence of any stellar
image of $m_{Ks} = 19.9$ mag at the 95\% confidence level.  Note that
the noise in the image is dominated by pixel-to-pixel background
variations which are repeatable between the individual subimages we
combined to make Figure 3.  This indicates that these variations are
due to background gradients from the crowded field rather than photon
statistics.

	Due to the poor seeing of the April 1994 data, only Stars 1-3
are detectable, and Stars 1 and 2 are badly blended, preventing
accurate photometry.  However, by comparison with nearby field stars,
we determine that Stars 1 and 3 did not vary by more than $\sim 0.8$
mag between April 1994 and June 1998.  Furthermore, the non-detection
of the other stars implies that they were all fainter than $K_s \sim
17.5$ mag in April 1994.

\subsection{GRS1758-258}

	In Figure 4, we cannot detect any sources inside our error
circle for GRS 1758-258.  We do detect 3 stars near this position
which were seen in I-band images by \citet{Marti98} -- labelled A, b,
and c.  Star A is the source identified as a candidate counterpart for
GRS 1758-258 by \citet{Marti98}, and for which they obtained an
infrared spectrum.  We present photometry of these 3 stars in Table 2,
following the same procedures described above for the field of
1E1740-2942.  Note that our $K_s$-band photometry of Star A is $\sim 2
\sigma$ brighter than the $K$-band photometry of \citet{Marti98}.
While the two filters do differ somewhat, this is a heavily reddened
star ($I-K = 5.0 $ mag, according to Marti et al.), so we would expect
the $K_s$-band flux to be {\it fainter} than the $K$-band flux.  Thus,
we have some marginal evidence for variability in Star A between
1994-1997 and June 1998.  At the exact center of our astrometric
solution for the position of GRS 1758-258, we can place an upper limit
on any stellar image of $m_{Ks} = 20.3$ mag at the 95\% confidence
level.

\section{Discussion}

\subsection{1E1740-2942}

	Our observations of 1E1740-2942 above essentially confirm those
of \citet{Marti00}.  We find virtually identical astrometric positions
for the microquasar on the infrared sky, with the same candidates
nearby, and very similar brightness for each of the candidates.  The
lack of apparent variability in the candidates, however, may indicate
that none of them are in fact associated with the microquasar, given
the known X-ray variability over the 1998-1999 time span
(e.g. \citet{Main}).

	The infrared extinction towards 1E1740-2942 is currently poorly
constrained, but based on the estimated X-ray column density of $N_H
\sim 0.5 - 1.0 \times 10^{23} \ {\rm cm^{-2}}$ \citep{chen94}, we
arrive at $A_{Ks} \sim 3.0 - 6.0$ mag.  This in turn allows us to
(very loosely) constrain the ranges of absolute magnitudes and thus
spectral types for the stars in this field (see Table 1).  As can be
seen in Table 1, all of the 8 candidate counterparts in the field are
consistent with both high-mass (early B or late O) main sequence
stars, as well as low-mass (K-type) giants, for example.  Thus, even
if we choose to associate one of these candidates with 1E1740-2942, we
unfortunately cannot place any interesting constraints on the nature
of the binary system.

\subsection{GRS1758-258}

	Our analysis of GRS 1758-258 presents two possible
conclusions, depending on which astrometric calibration of its
position is correct -- the one we present above, or that of
\citet{Marti98}.  In the latter case, then Star A or one of the two
stars near the ``M'' in Figure 4 is likely to be the IR counterpart to
GRS 1758-258.  Assuming $A_{Ks} = 0.9$ mag based on $N_H = 1.5 \pm 0.1
\times 10^{22} {\ \rm cm^{-2}}$ from \citet{Mereghetti}, and a
distance of 8.5 kpc, we derive absolute magnitudes for the three stars
of $K_s = -1.9, +1.35, +1.20$ mag, respectively.  These are roughly
consistent with an early K-type giant for Star A (as proposed by
\citet{Marti98}), and early A-type main sequence stars for the other
two stars.  Any of these is inconsistent with a high-mass binary
companion scenario for GRS 1758-258, and instead indicates that the
system is an intermediate- or low-mass binary.

	However, as noted above, our own astrometric solution would
rule out all three of these stars as counterparts to GRS 1758-258.
Since our solution is based on and consistent with the astrometric
positioning of 12 USNO field stars with small internal scatter, the
only possibility which would explain such a discrepancy is that the
optical/radio frame tie is inaccurate for this region of sky at the
$\sim 1 \arcsec$ level.  Since the USNO frame is tied to the
International Coordinate Reference Frame via HIPPARCOS, with a claimed
accuracy of $<1$ mas \citep{Perryman}, this seems to be unlikely.
Thus, we conclude that the error circle in Figure 4 is in fact the
most accurate positional determination for GRS 1758-258.

	In that case, the lack of an IR counterpart to $m_{Ks} > 20.6$
mag is particularly intriguing.  Taking the same extinction and
distance values as above, we place a limit on the absolute magnitude
of any counterpart of $K_s > +4.9$ mag.  This rules out any companion
more luminous than a K7 main-sequence star.  If this is correct, then
GRS 1758-258 must be a (very) low-mass X-ray binary.

	As a final alternative, it may be that the radio position is
for some reason incorrect at the $\sim 1-2$ arcsecond level, which
would leave several candidates for IR counterparts to GRS 1758-258.

\section{Conclusions}

	We have presented deep infrared ($2.2 \mu$m) imaging of the
fields of the Galactic microquasars 1E1740-2942 and GRS 1758-258, along
with astrometric positioning to the $\sim 0.5-0.6 \arcsec$ level.  We
summarize our results as follows:

\begin{itemize}

\item For 1E1740-2942, our astrometric position agrees with that of
\citet{Marti00}.

\item We find all 8 candidate counterparts to 1E1740-2942 identified by
\citet{Marti00}, and no new candidates.

\item None of the candidate IR counterparts show significant
variability from June 1998 to May 1999, while the X-rays from
1E1740-2942 {\it do} show variability over that epoch.

\item For GRS 1758-258, our astrometric position, based on 12 nearby
USNO reference stars, disagrees with that of \citet{Marti98} by $\sim
1.5 \arcsec$.

\item We find no IR counterpart inside our 90\% confidence level error
circle down to a $3 \sigma$ limit of $m_{Ks} > 20.3$ mag.  For assumed
$A_{Ks} = 0.9$ mag and $d = 8.5$ kpc, this limits any companion to be
fainter than a main-sequence K5 star, implying that GRS 1758-258 is a
low-mass binary.

\item Alternately, if the astrometric position of \citet{Marti98} is
correct, we see only marginal evidence for variability from the
brightest star in their error circle between June 1998 and the
1994-1997 photometry of \citet{Marti98}.

\end{itemize}

\acknowledgments The authors thank L. Armus and R. Gal for help in
obtaining the CCD images at Palomar Observatory, and M. Pahre for help
in obtaining the 1994 Keck data.  We also thank the staffs of Palomar
and Keck Observatories for their expert help during our observing
runs.  SSE is supported in part at Cornell by an NSF CAREER award
(NSF-9983830) and was supported at Caltech by the Sherman Fairchild
Postdoctoral Fellowship in Physics.  WJF was supported at Cornell by
the NSF REU summer fellowship program.  SGD acknowledges partial
support from the Bressler Foundation.

\vfill \eject

\begin{deluxetable}{cccl}
\tablecaption{Photometry of Objects Near 1E1740-2942}
\startdata
Star Number & $m_{Ks}$ (mag) & $K_s \ \rm{(mag)}$\tablenotemark{a} & Spectral Type\tablenotemark{b}\\
\hline 
1 & $17.1 \pm 0.1$ & $-$0.2 -- $-$3.4 & B7V-B0V , K5III-G0III \\
2 & $18.2 \pm 0.1$ & +0.9 -- $-$2.3 & A1V-B1V , K2III-G0III \\
3 & $16.8 \pm 0.1$ & $-$0.5 -- $-$3.7 & B6V-O9V , K5III-G0III \\
4 & $18.2 \pm 0.1$ & +0.9 -- $-$2.3 & A1V-B1V , K2III-G0III \\
5 & $18.7 \pm 0.2$ & +1.4 -- $-$1.8 & A2V-B2V , K1III-G0III \\
6 & $19.1 \pm 0.2$ & +1.8 -- $-$1.4 & A6V-B3V , G8III-G0III \\
7 & $19.1 \pm 0.2$ & +1.8 -- $-$1.4 & A6V-B3V , G8III-G0III \\
8 & $19.5 \pm 1.0$ & +3.1 -- $-$1.9 & G0V-B2V , K1III-G0III \\
\enddata
\tablenotetext{a}{Assuming $A_{Ks} = 3-6 $ mag and $d=8.5 $ kpc.}
\tablenotetext{b}{Ranges given assuming luminosity class of III or V.  All stars of luminosity class I are excluded.}
\end{deluxetable}

\begin{deluxetable}{cccl}
\tablecaption{Photometry of Objects Near GRS 1758-258}
\startdata
Star Number & $m_{Ks}$ (mag) & $K_s \ \rm{(mag)}$\tablenotemark{a} & Spectral Type \\
\hline 
A & $13.63 \pm 0.13$ & $-$1.90 & K0III \\
b & $16.73 \pm 0.16$ & +1.35 & AOV \\
c & $16.88 \pm 0.16$ & +1.20 & AOV \\
\enddata
\tablenotetext{a}{Assuming $A_{Ks} = 0.9 $ mag and $d=8.5 $ kpc.}
\end{deluxetable}

\begin{figure}
\plotone{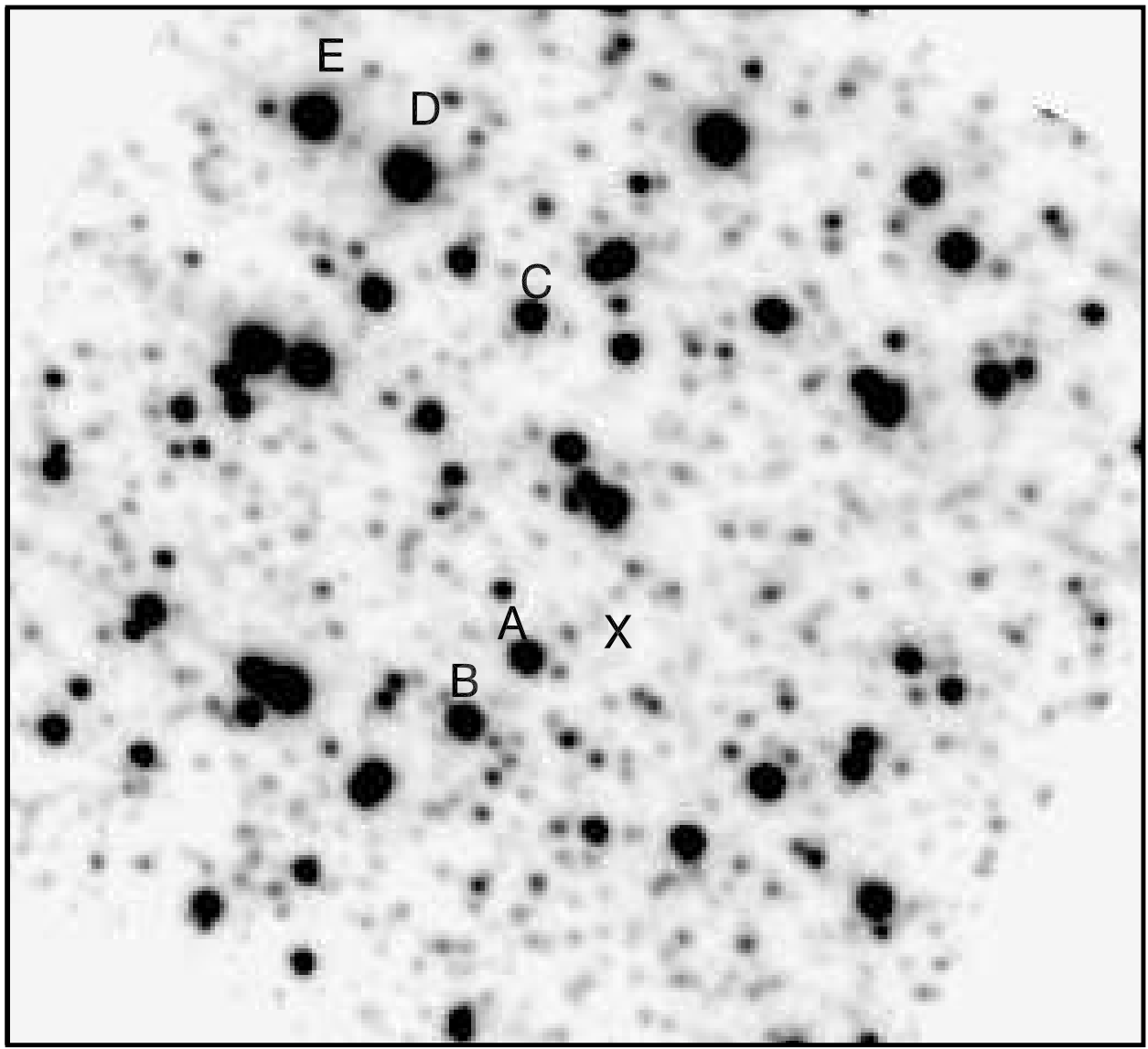}
\caption{\it Keck $K_s$-band image of the field of GRS 1758-258
(approximately $1 \arcmin$ on a side).  ``X'' indicates the position
of the microquasar.  Letters indicates the stars visible in both the
K-band and I-band CCD image used for astrometric calibration.  North
is up and east is to the left.}
\end{figure}

\begin{figure}
\plotone{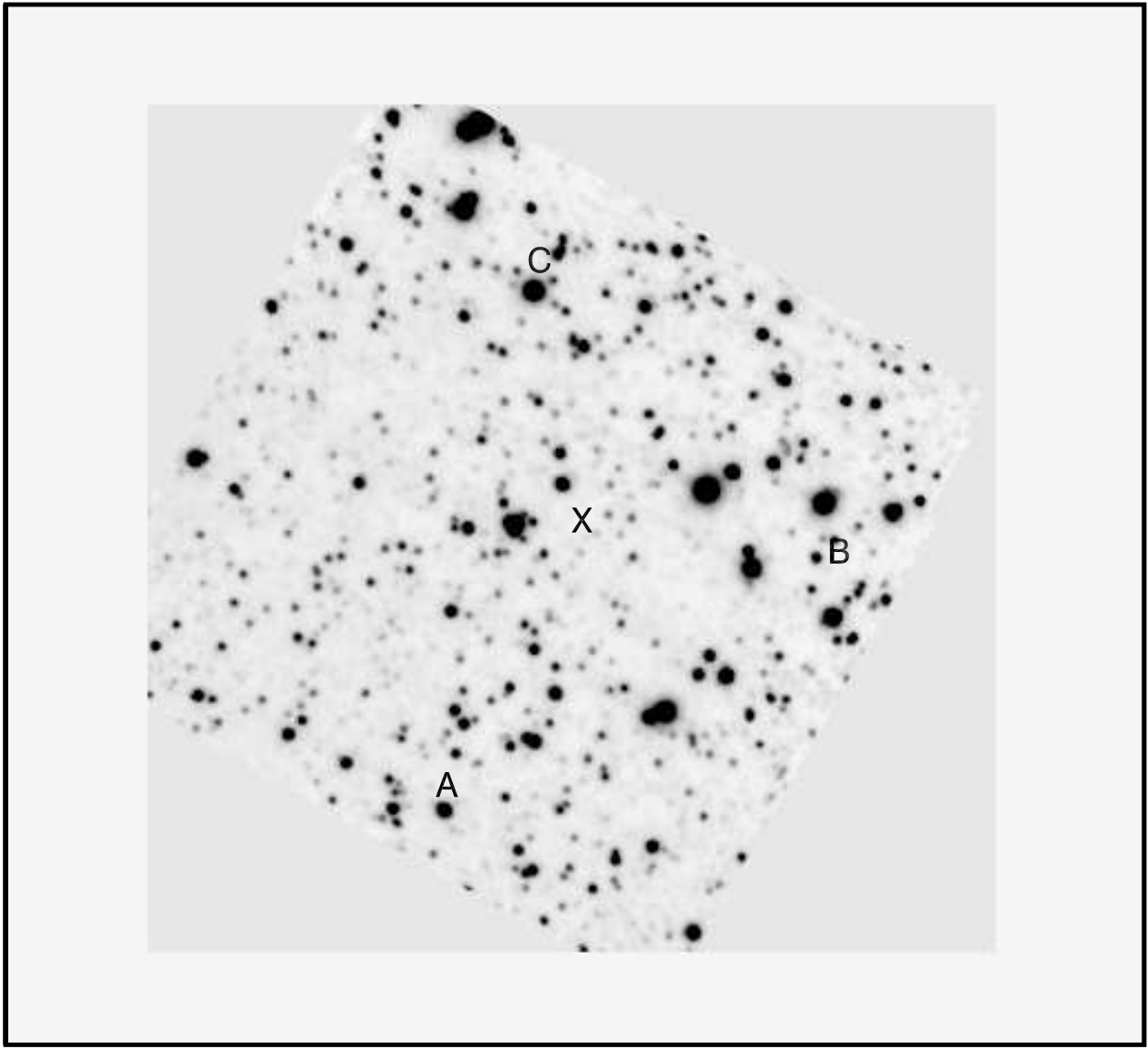}
\caption{\it Keck $K_s$-band image of the field of 1E1740-2942
(approximately $1 \arcmin$ on a side).  ``X'' indicates the position
of the microquasar.  ``A'', ``B'', and ``C'' are the USNO stars used
for astrometric calibration of the image (``C'' is the brightest star
near the symbol).  North is up and east is to the left.}
\end{figure}

\begin{figure}
\plottwo{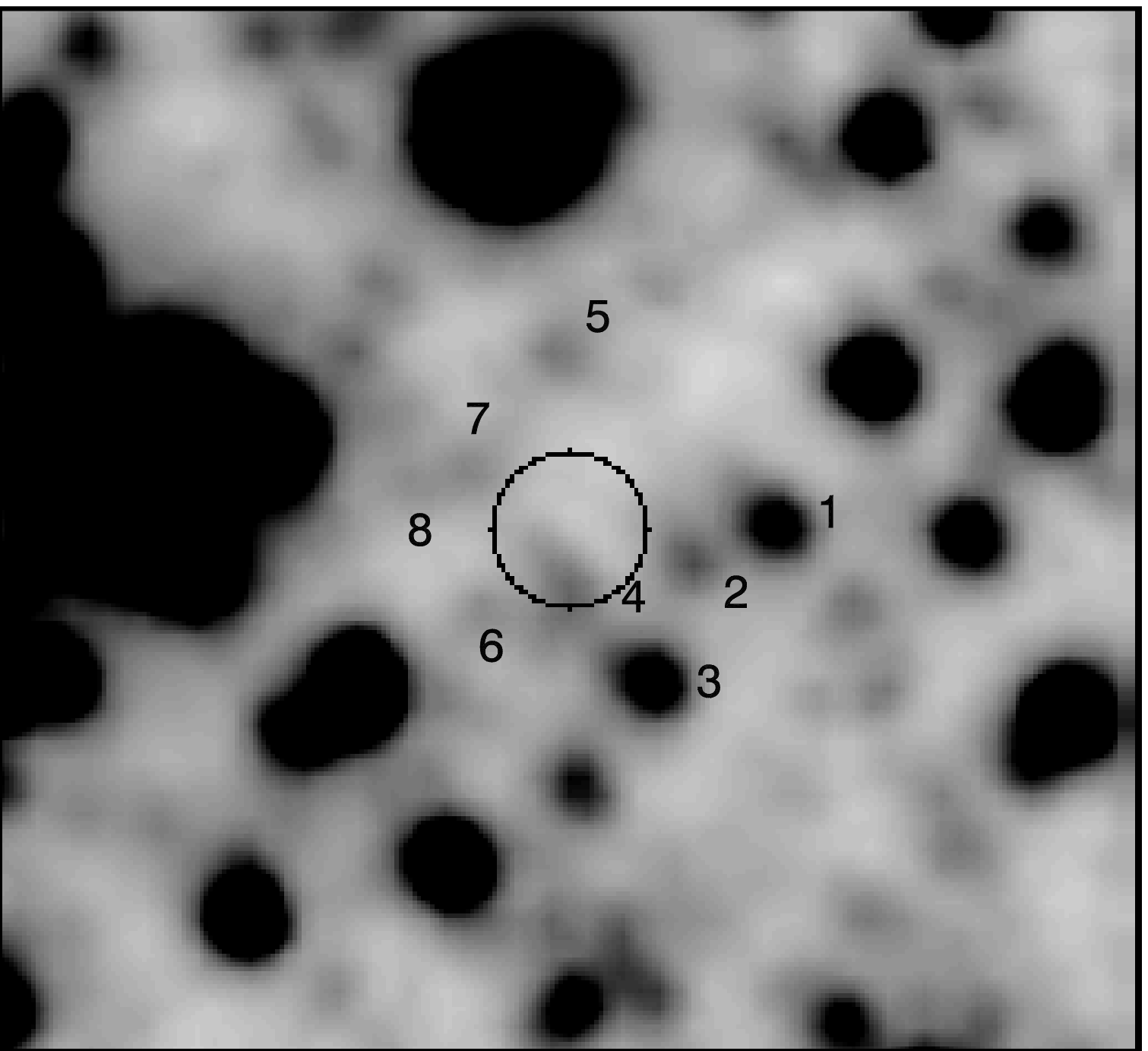}{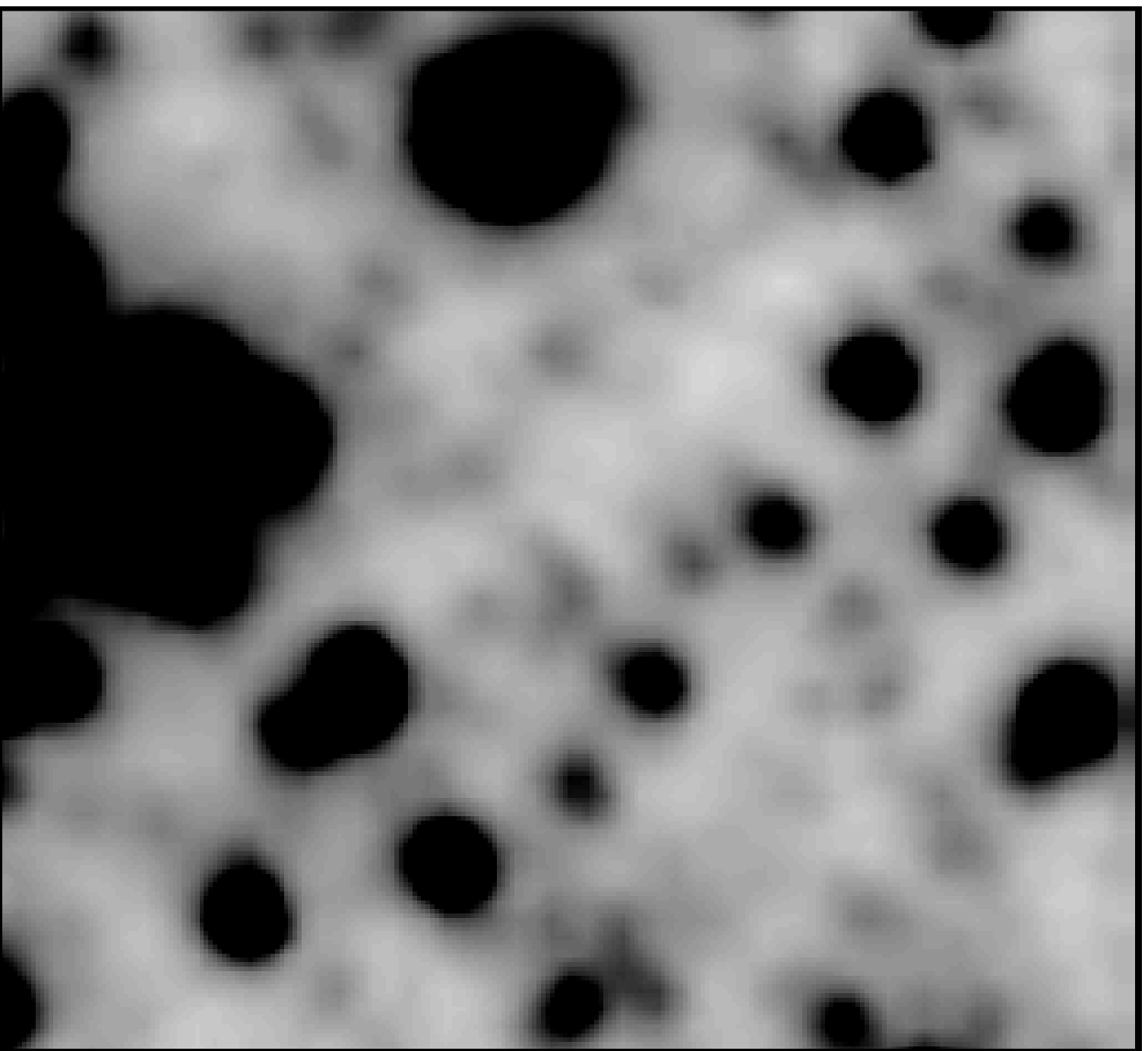}
\caption{\it Close-up ($10 \arcsec \times 10 \arcsec$) Keck $K_s$-band
image of the field of 1E1740-2942.  The left side image shows the 90\%
error circle ($1.3 \arcsec$ diameter) for the location of the
microquasar, with numbers indicating the potential candidates reported
by \citet{Marti00}.  The right hand side is an identical copy of
the image displayed without markings, for clarity.}
\end{figure}

\begin{figure}
\plotone{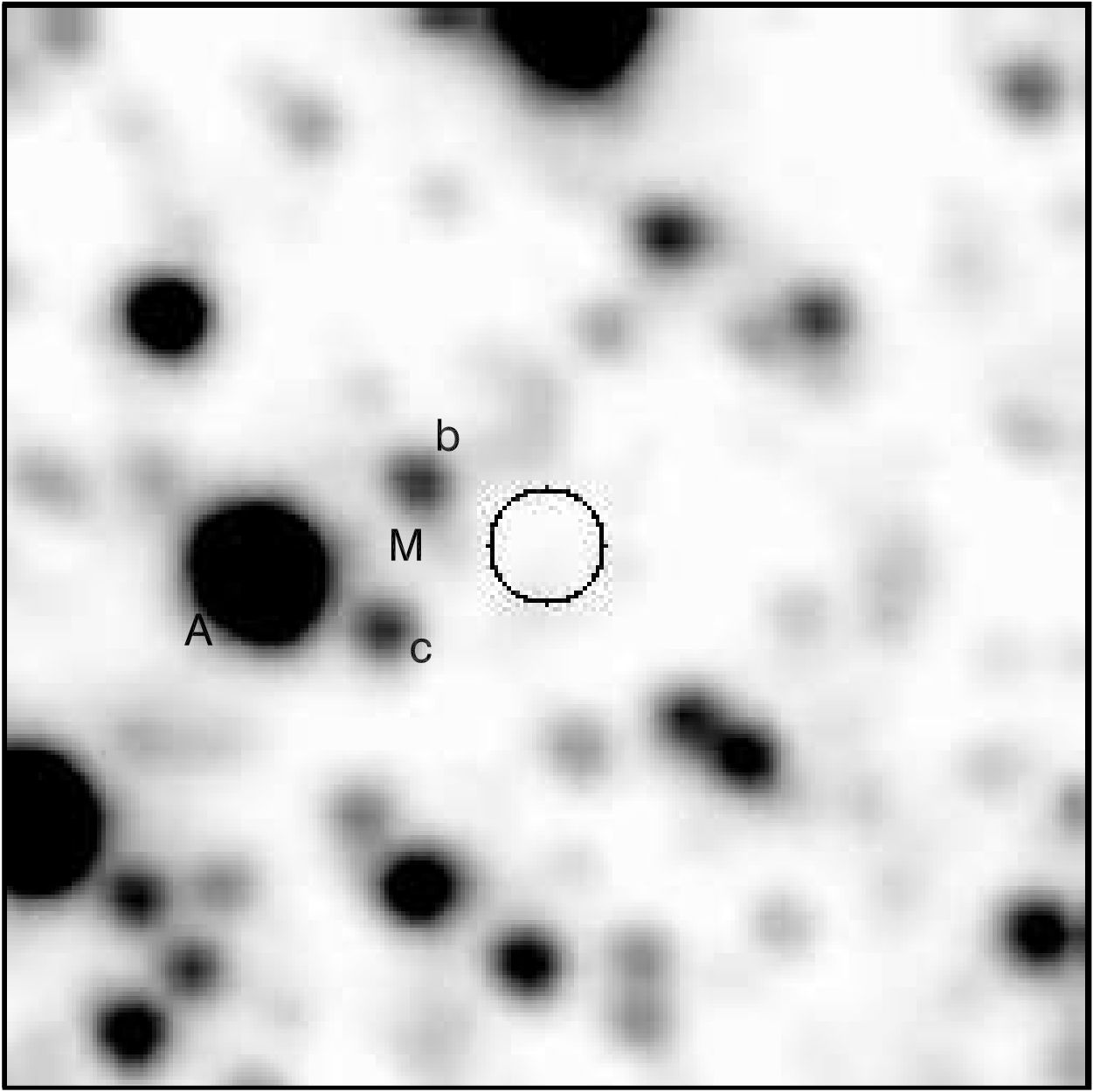}
\caption{\it Close-up ($10 \arcsec \times 10 \arcsec$) Keck $K_s$-band
image of the field of GRS 1758-258, including the 90\% error circle
($1.0 \arcsec$ diameter) for the microquasar position.  ``M''
indicates the approximate position for the microquasar determined by
\citet{Marti98}.}
\end{figure}


\begin{thebibliography}

\bibitem[Chen et al., 1994]{chen94} Chen, Wan, Gehrels, Neil, Leventhal, Marvin 1994, \apj, 426, 586

\bibitem[da Silva Neto, et al., 2000]{daSilva} da Silva Neto, Dario
N., Andrei, A. H., Martins, R. Vieira, Assafin, M. 2000, \aj, 119, 1470

\bibitem[Djorgovski et al., 1992]{George} Djorgovski, S., Thompson, D., Mazzarella, J., Klemola, A., Neugebauer,
 G., Matthews, K., Armus, L. 1992, IAU Circular 5596

\bibitem[Kuznetsov et al., 1996]{Kuz} Kuznetsov, S., Gilfanov, M.,
Churazov, E., Sunyaev, R., Vikhlinin, A., Khavenson, N., Dyachkov, A.,
Laurent, P., Goldwurm, A., Cordier, B., Vargas, M., Mandrou, P.,
Roques, J. P., Jourdain, E., Borrel, V. 1996, Proc. 'Roentgenstrahlung
from the Universe', eds. Zimmermann, H.U., Trümper, J., and Yorke, H.,
MPE Report 263, p. 157-158

\bibitem[Main et al., 1999]{Main} Main, D. S., Smith, D. M., Heindl,
W. A., Swank, J., Leventhal, M., Mirabel, I. F., Rodríguez,
L. F. 1999, \apj, 525, 901

\bibitem[Marti et al., 1998]{Marti98} Marti, J., Mereghetti, S.,
Chaty, S., Mirabel, I. F., Goldoni, P., Rodr\'\i guez, L. F 1998, \aap,
338, L95

\bibitem[Marti et al., 2000]{Marti00} Marti, J., Mirabel, I. F.,
Chaty, S., Rodríguez, L. F., 2000, \aap, 363, 184

\bibitem[Matthews \& Soifer, 1994]{NIRC} Matthews, K., Soifer,
B. T. 1994, ExA, 3, 77

\bibitem[Mereghetti et al., 1997]{Mereghetti} Mereghetti, S.,
Cremonesi, D. I., Haardt, F., Murakami, T., Belloni, T., Goldwurm,
A. 1997, \apj, 476, 829

\bibitem[Mirabel et al., 1991]{Felix91} Mirabel, I. F., Paul, J.,
Cordier, B., Morris, M., Wink, J. 1991, \aap, 251, L43

\bibitem[Mirabel et al., 1992]{Felix92} Mirabel, I. F., Rodr\'\i guez,
L. F., Cordier, B., Paul, J., Lebrun, F. 1992, \nat, 358, 215

\bibitem[Mirabel \& Rodr\'\i guez, 1993]{Felix93} Mirabel, I. F.,
Rodr\'\i guez, L. F., 1993, AIP Conference Proceedings 304, The Second
Compton Symposium, 413

\bibitem[Mirabel \& Rodr\'\i guez, 1994]{Felix94} Mirabel, I. F.,
Rodr\'\i guez, L. F., 1994, \nat, 371, 46

\bibitem[Monet et al., 1998]{USNO} Monet D.  Bird A., Canzian B., Dahn
C., Guetter H., Harris H., Henden A., Levine S., Luginbuhl C., Monet
A.K.B., Rhodes A., Riepe B., Sell S., Stone R., Vrba F., Walker
R. 1998, USNO-A V2.0, A Catalog of Astrometric Standards

\bibitem[Perryman et al., 1997]{Perryman} Perryman M.A.C., Lindegren
L., Kovalevsky J., Hog E., Bastian U., Bernacca P.L., Creze M., Donati
F., Grenon M., Grewing M., van Leeuwen F., van der Marel H., Mignard
F., Murray C.A., Le Poole R.S., Schrijver H., Turon C., Arenou F.,
Froeschle M., Petersen C.S. 1997, \aap, 323, L49

\bibitem[Rodr\'\i guez et al., 1992]{Luis} Rodr\'\i guez, L. F., Mirabel,
I. F., Marti, J., 1992, \apj, 401, L15

\bibitem[Sunyaev et al., 1991]{Sunyaev} Sunyaev, R., Churazov, E.,
Gilfanov, M., Pavlinsky, M., Grebenev, S., Babalyan, G., Dekhanov, I.,
Khavenson, N., Bouchet, L., Mandrou, P., Roques, J. P., Vedrenne, G.,
Cordier, B., Goldwurm, A., Lebrun, F., Paul, J 1991, \apj, 383, L49

\bibitem[Tingay et al., 1995]{Tingay}
Tingay, S. J., Jauncey, D. L., Preston, R. A., Reynolds, J. E., Meier,
 D. L., Murphy, D. W., Tzioumis, A. K., McKay, D. J., Kesteven, M. J.,
 Lovell, J. E. J., Campbell-Wilson, D., Ellingsen, S. P., Gough, R.,
 Hunstead, R. W., Jones, D. L., McCulloch, P. M., Migenes, V., Quick,
 J., Sinclair, M. W., Smits, D. 1995, \nat, 374, 141

\end{thebibliography}
\end{document}